# Influence of Dimensionality on Thermoelectric Device Performance

Raseong Kim, Supriyo Datta, and Mark S. Lundstrom

Network for Computational Nanotechnology

Discovery Park, Purdue University, West Lafayette, IN 47907

**Abstract**

The role of dimensionality on the electronic performance of thermoelectric devices is clarified using the Landauer formalism, which shows that the thermoelectric coefficients are related to the transmission, $T(E)$, and how the conducting channels, $M(E)$, are distributed in energy. The Landauer formalism applies from the ballistic to diffusive limits and provides a clear way to compare performance in different dimensions. It also provides a physical interpretation of the "transport distribution," a quantity that arises in the Boltzmann transport equation approach. Quantitative comparison of thermoelectric coefficients in one, two, and three dimension shows that the channels may be utilized more effectively in lower-dimensions. To realize the advantage of lower dimensionality, however, the packing density must be very high, so the thicknesses of the quantum wells or wires must be small. The potential benefits of engineering $M(E)$ into a delta-function are also investigated. When compared to a bulk semiconductor, we find the potential for ~50 % improvement in performance. The shape of $M(E)$ improves as dimensionality decreases, but lower dimensionality itself does not guarantee better performance because it is controlled by both the shape and the magnitude of $M(E)$. The benefits of engineering the shape of $M(E)$ appear to be modest, but approaches to increase the magnitude of $M(E)$ could pay large dividends.

## 1. Introduction

The efficiency of thermoelectric devices is related to the figure of merit, $ZT = S^2GT/\kappa$,[1] where $T$ is the temperature, $S$ is the Seebeck coefficient, $G$ is the electrical conductance, and $\kappa$ is the thermal conductance, which is the sum of the electronic contribution, $\kappa_e$, and the lattice thermal conductance, $\kappa_l$. The use of artificially structured materials such as superlattices[2] and nanowires[3,4] has proven to be an effective way to increase the performance of thermoelectric devices by suppressing phonon transport. In addition to the success of phonon engineering, additional benefits might be possible by enhancing the electronic performance of thermoelectric devices.[5] Possibilities include reducing device dimensionality[6,7] and engineering the bandstructure.[5]

Using the Boltzmann transport equation (BTE),[8] thermoelectric transport coefficients can be expressed in terms of the "transport distribution," $\Xi(E)$.[9,10] Note that the quantity, $q^2\Xi(E)$, is sometimes called "differential conductivity," $\sigma(E)$,[11,12] where $q$ is the electron charge. Mahan and Sofo[9] showed mathematically that a delta-shaped $\Xi(E)$ gives the best thermoelectric efficiency. It has been also shown that the efficiency approaches the Carnot limit for a delta-shaped $\Xi(E)$ when the phonon heat conduction tends to zero.[13]

An alternative approach, the Landauer formalism,[14] has been widely-used in mesoscopic thermoelectric studies.[15-18] In this paper, we show that it is also useful for macroscopic thermoelectrics. The Landauer formalism reduces to the diffusive results that can be also obtained from the BTE for large structures and to the ballistic results for small structures. It also provides a useful physical interpretation of conventional results from the BTE and a convenient way to compare performance across dimensions.

It has been reported that one-dimensional (1D) and two-dimensional (2D) structures may provide enhanced electronic performance due to the increased electrical conductivity per unit volume.[6,7,19] Also, it has been argued that 1D thermoelectric devices will give better efficiencies because the density-of-states is close to a delta-function.[12] Comparisons across dimensions, however, are not straightforward due to the issues such as the assumed cross section of nanowires and their packing density in a three-dimensional (3D) structure.[20]

In this paper, our objective is to examine the role of dimensionality on the electronic performance of thermoelectric devices using the Landauer formalism. Similar comparisons have been done in the past,[6,7] but comparisons across dimensions are often clouded by assumptions about the nanowire diameter and packing fraction. We present an approach that bypasses these issues. The paper is organized as follows. In Section 2, we summarize the Landauer approach and present a physical interpretation of $\Xi(E)$, which turns out to be proportional to the transmission function, $\bar{T}(E)$.[21] In Section 3, we compare the Seebeck coefficient, $S$, and power factor ($S^2G$) in 1D, 2D, and 3D ballistic devices and discuss the role of dimensionality. In Section 4, scattering is briefly discussed, and we examine the upper limit of performance possible by shaping $\bar{T}(E)$ into a delta-function. Conclusions follow in Section 5.

## 2. Approach

According to the Landauer formalism,[14] the electrical current (*I*) and heat current are expressed as

$$I = \frac{2q}{h} \int T(E) M(E) (f_1 - f_2) dE \qquad [A] \qquad (1)$$

$$I_{q1} = \frac{2}{h}\int T(E)M(E)(E-E_{F1})(f_1-f_2)dE \qquad [W] \qquad (2a)$$

$$I_{q2} = \frac{2}{h}\int T(E)M(E)(E-E_{F2})(f_1-f_2)dE \qquad [W] \qquad (2b)$$

where $I_{q1}$ and $I_{q2}$ are the heating and cooling rates of contact 1 and contact 2 respectively, and $I_{q2} - I_{q1} = \Delta VI$ where $\Delta V$ is the voltage difference between the contacts.[22] In (1)-(2), $h$ is the Planck constant, $T(E)$ is the transmission, $M(E)$ is the number of conducting channels at energy, $E$, $E_{F1}$ and $E_{F2}$ are the Fermi levels of the two contacts, and $f_1$ and $f_2$ are equilibrium Fermi-Dirac distributions for the contacts. In this paper, we assume a uniform conductor in which $T(E)$ is determined by scattering. Equations (1) and (2) apply to ballistic devices (commonly referred to as thermionic devices) as well as to diffusive devices (commonly referred to as thermoelectric devices). For ballistic devices, $T(E) = 1$, and for diffusive devices, $T(E) = \lambda(E)/(\lambda(E)+L) \approx \lambda(E)/L$, where $\lambda(E)$ is the energy-dependent mean-free-path and $L$ is the length of the conductor.[23]

In the linear response regime, $I_{q2} \approx I_{q1} \equiv I_q$, and (1) and (2) are expressed as

$$I = \langle G \rangle \Delta V + \langle SG \rangle \Delta T \qquad (3)$$

$$I_q = -T\langle SG \rangle \Delta V - \kappa_0 \Delta T \qquad (4)$$

where $\Delta T$ is the temperature difference between contacts. We set $E_F = E_{F1}$ and $f = f_1$, and the transport coefficients are

$$\langle G \rangle = \frac{2q^2}{h}\int_{-\infty}^{\infty} \bar{T}(E)\left(-\frac{\partial f}{\partial E}\right)dE \qquad [1/\Omega] \qquad (5)$$

$$T\langle GS \rangle = -\frac{2q}{h} \int_{-\infty}^{\infty} \bar{T}(E)(E-E_F)\left(-\frac{\partial f}{\partial E}\right) dE \quad [\text{V}/\Omega] \quad (6)$$

$$\kappa_0 = \frac{2}{hT} \int_{-\infty}^{\infty} \bar{T}(E)(E-E_F)^2 \left(-\frac{\partial f}{\partial E}\right) dE \quad [\text{W/K}] \quad (7)$$

where $\bar{T}(E)$ is the transmission function,[21] $\bar{T}(E) = T(E)M(E)$, and $\kappa_0$ is the electronic thermal conductance for zero electric field. Note that the units indicated in (5)-(7) are the same in all three dimensions. Alternatively, (3) and (4) can be expressed as

$$\Delta V = I/G - S\Delta T \quad (8)$$

$$I_q = \Pi I - \kappa_e \Delta T \quad (9)$$

where $G = \langle G \rangle$, $S = \langle SG \rangle / \langle G \rangle$, $\Pi$ is the Peltier coefficient, $\Pi = TS$, and $\kappa_e = \kappa_0 - TS^2 G$.

Comparing the transport coefficients in (5)-(7) with those from the BTE,[9] we observe that the "transport distribution" $\Xi(E)$[9,10] has a simple, physical interpretation; it is proportional to $\bar{T}(E)$ as

$$\Xi(E) = \frac{2}{h}\bar{T}(E) = \frac{2}{h}T(E)M(E), \quad (10)$$

where $M(E)$ essentially corresponds to the carrier velocity times the density-of-states,[21] and $T(E)$ is a number between zero and one that is controlled by carrier scattering. (Note that $T(E)$ can also be engineered in quantum structures such as superlattices,[24,25] a possibility not considered in this paper.)

In this section, we assume ballistic conductors with $T(E) = 1$. As noted earlier, the expressions for transport coefficients as given by (5)-(7) are the same for all dimensions; only $M(E)$ changes. In this study, we will assume a simple energy bandstructure (although we believe that the overall conclusions are more general). If we assume that a single parabolic subband is occupied,

$$M_{1D}(E) = \Theta(E - \varepsilon_1) \quad (11a)$$

$$M_{2D}(E) = W \frac{\sqrt{2m^*(E - \varepsilon_1)}}{\pi \hbar} \quad (11b)$$

$$M_{3D}(E) = A \frac{m^*}{2\pi \hbar^2}(E - E_C) \quad (11c)$$

where $\Theta$ is the unit step function, $\hbar = h/2\pi$, $\varepsilon_1$ is the bottom of the first subband, $m^*$ is the electron effective mass, $E_C$ is the conduction band edge, and $W$ and $A$ are the width and the area of the 2D and 3D conductors, respectively. Sketches in Fig.1 clearly show the difference between the density-of-states and $M(E)$ for 1D, 2D, and 3D conductors. Using (5)-(7) and (11), thermoelectric transport coefficients can be calculated and compared across dimensions as discussed in the following section.

## 3. Results

In this section, we compare each component of $ZT$ determined by electronic properties for 1D, 2D, and 3D ballistic conductors. Seebeck coefficients can be compared across dimensions directly because they have the same units in all dimensions. They are calculated from

$$S = \left(\frac{k_B}{-q}\right) \frac{\int M(E)\left[(E - E_F)/k_B T\right]\left(-\frac{\partial f}{\partial E}\right) dE}{\int M(E)\left(-\frac{\partial f}{\partial E}\right) dE} \quad [\text{V/K}] \quad (12)$$

where $k_B$ is the Boltzmann constant. For the following model calculations, we assume $T = 300$ K and $m^* = m_0$, where $m_0$ is the free electron mass. Fig. 2 plots $S$ vs. $\eta_F$ for 1D, 2D, and 3D ballistic conductors, where $\eta_F$ is the position of $E_F$ relative to the band edge, $\eta_F = (E_F - \varepsilon)/k_B T$. (To first order, $S$ is independent of scattering as discussed in Section 4.) Fig. 2 shows that $|S_{3D}| > |S_{2D}| > |S_{1D}|$ for the same $\eta_F$. As shown in (12), $S$ increases as the separation between $E_F$ and $M(E)$ increases. As the dimensionality increases, $M(E)$ in (11) spreads out more, so $S$ improves.

Although Fig. 2 shows that the magnitude of $S$ is greater in 3D than in 1D or 2D for any $\eta_F$, there is more to the story. The power factor, $S^2 G$, is an important part of the $ZT$. As shown in Fig. 3, the power factor displays a maximum at $\eta_{F,\max} = -1.14, -0.367$, and $0.668$ in 1D, 2D, and 3D respectively. This occurs because the electrical conductance,

$$G = \frac{2q^2}{h} \int M(E) \left( -\frac{\partial f}{\partial E} \right) dE \equiv \frac{2q^2}{h} M_{eff} \qquad [1/\Omega], \qquad (13)$$

where $M_{eff}$ is the effective number of conducting channels, increases more rapidly with $\eta_F$ in 1D than in 2D or 3D as shown in Fig. 4. If we compare $S$ not at the same $\eta_F$ but rather at the $\eta_{F,\max}$ in each dimension, then $\hat{S} = |S(\eta_{F,\max})|$ is highest in 1D and lowest in 3D as indicated by the arrows in Fig. 2. For the specific case considered, $\hat{S}_{1D} = 2.29 \times k_B/q$, $\hat{S}_{2D} = 2.16 \times k_B/q$, and $\hat{S}_{3D} = 1.94 \times k_B/q$, which shows an 18 % improvement in 1D over 3D.

The power factor is a key figure of merit for thermoelectric devices, but comparing power factors across dimensions brings up issues of the size and packing densities of the nanowires or quantum wells[20] because $G$ is proportional to $M_{eff}$, which depends on $W$ and $A$ for 2D and 3D conductors respectively. An alternative approach is to compare the power factor per mode, $S^2 G / M_{eff}$, at $\eta_{F,\max}$ for each dimensionality. The quantity, $S^2 G / M_{eff}$, has the units $[W/K^2]$ and can therefore be compared directly across dimensions. The results are $S^2G/M_{eff}\big|_{1D} = 5.24 \times 2k_B^2/h$, $S^2G/M_{eff}\big|_{2D} = 4.68 \times 2k_B^2/h$, and $S^2G/M_{eff}\big|_{3D} = 3.75 \times 2k_B^2/h$. We observe that the modes are more effectively used in 1D and 2D than in 3D. In 1D, the power factor per mode is 40 % larger than in 3D and 12 % larger than in 2D. The benefits come from the fact that $\hat{S}$ is highest in 1D and lowest in 3D.

So far, we have demonstrated that 1D thermoelectrics are superior to 3D thermoelectrics in terms of the Seebeck coefficient at the maximum power factor and in terms of the power factor per mode. To make use of 1D thermoelectric devices in macroscale applications, many nanowires must be placed in parallel, so issues of the nanowire size and packing density arise. To illustrate the considerations involved, we present a simple example. We first compute the maximum power factor for a 3D device ($S^2G_{3D,max}$) with an area of 1 cm$^2$. For our model device with ballistic conduction, the result is $S^2G_{3D,max} = 12.6$ W/K$^2$. We also find the number of effective conducting channels from (13) as $M_{eff,3D} = 5.84 \times 10^{12}$. To compare this performance to a 2D thermoelectric device, we compute the maximum power factor of a 2D device ($S^2G_{2D,max}$) with $W$ = 1 cm, and the model calculation shows that $S^2G_{2D,max} = 2.96 \times 10^{-6}$ W/K$^2$ and $M_{eff,2D} = 1.10 \times 10^6$. Finally, we do the same for a 1D device and find $S^2G_{1D,max} = 7.28 \times 10^{-13}$ W/K$^2$ for $M_{eff,1D} = 0.242$.

The analysis presented earlier established that the power factor per mode is significantly better in 1D than in 2D, which is, in turn better than 3D. To realize this advantage on the 1 cm × 1cm scale, we must produce the same number of effective modes in that area as is achieved in 3D. To do so (assuming 100% packing fraction), we find that the thickness of the 2D films must be less than $M_{eff,2D}/M_{eff,3D}$ ~ 1.89 nm or the size of each nanowire must be less than $\left(M_{eff,1D}/M_{eff,3D}\right)^{1/2} \times \left(M_{eff,1D}/M_{eff,3D}\right)^{1/2}$ ~ 2.03×2.03 nm. Alternatively, we could seek to achieve the same power factor and ask what the size of the thin film or nanowire would need to be (still assuming a 100% packing fraction). The answer is 2D films with a thickness of $S^2 G_{3D,max}/S^2 G_{2D,max}$ ~ 2.35 nm or 1D nanowires with a size of $\left(S^2 G_{1D,max}/S^2 G_{3D,max}\right)^{1/2} \times \left(S^2 G_{1D,max}/S^2 G_{3D,max}\right)^{1/2}$ ~ 2.40×2.40 nm. However we choose to look at it, the conclusion is that to realize the benefits of the inherently better thermoelectrics performance in 2D or 1D requires very small structures with very high packing fractions. In practice, nanowires or quantum wells should be separated by barriers,[19] i.e. the "fill factor" should be less than 1. For wires, the fill factor should be 1/4 to 1/3 to maintain their 1D properties.[20] This means that the advantages coming from the more nearly optimal distribution of modes for 1D systems are likely to be compensated by the limited fill factor, which reduces the total number of modes, $M_{eff}$.

Finally, we should consider how different $m^*$ might affect our conclusions. To first order, $S$ is independent of $m^*$. In 1D, $G$ is also independent of $m^*$ because the $m^*$-dependencies in the density-of-states and the velocity cancel out.[26] In 2D $G_{2D} \propto \sqrt{m^*_{2D}}$, and in 3D $G_{3D} \propto m^*_{3D}$. Re-doing the analysis presented above for $m^* = 0.1 m_0$, we find that the required sizes of the 2D films or 1D wires is about three times larger. Although the size and packing fraction requirements are still daunting, it appears that the use of low-dimensional structures may be more advantageous for low effective mass thermoelectric devices

## 4. Discussion

Our analysis so far has assumed ballistic transport, $T(E) = 1$; in the diffusive limit, $T(E) \to \lambda(E)/L$. For several common scattering mechanisms, $\lambda(E)$ can be expressed in power law form as $\lambda(E) = \lambda_0 (E(p)/k_B T)^s$, where $\lambda_0$ is a constant, $E(p)$ is the kinetic energy, and $s$ is the characteristic exponent, which depends on device dimensionality and the particular scattering mechanisms.[27] Using this form, the transport coefficients for diffusive thermoelectrics can be calculated from (5)-(7). We compare the power factor per mode, $S^2 G/M_{eff}$, for three cases: i) an energy-independent $\lambda(E)$ with $s = 0$, ii) a constant scattering time ($\tau$) with $s = 1/2$, and iii) scattering rates ($1/\tau$) proportional to the density-of-states, where $s$ is 1, 1/2, and 0 for 1D, 2D, and 3D respectively. In case i), the results are the same as the ballistic case because $S$ does not depend on scattering and $G$ is simply scaled by a factor of $\lambda_0/L$. In case ii), the modes are still utilized more effectively in lower dimensions as $S^2 G/M_{eff}\big|_{1D} = 4.68 \times 2k_B^2/h$, $S^2 G/M_{eff}\big|_{2D} = 3.75 \times 2k_B^2/h$, and $S^2 G/M_{eff}\big|_{3D} = 2.26 \times 2k_B^2/h$. In this case, the 40 % improvement that we found in the ballistic case has become a 100 % improvement of 1D over 3D. In case iii), however, the power factor per mode is the same in all dimensions, $S^2 G/M_{eff} = 3.75 \times 2k_B^2/h$. A full treatment of the role of scattering is beyond the scope of this study. It involves more than the characteristic exponent because effects such as surface roughness scattering[28,29], enhanced phonon scattering[30], and interaction with confined phonon modes[31] may arise in 1D structures. From the solutions of the inelastic 3D Boltzmann equation, Broido and Reinecke[32] have shown that there is a limit to the enhancement of the power factor in the quantum well and quantum wire because scattering rates increase with decreasing well and wire widths. The comparison of the electronic and lattice contributions to *ZT* for specific III-V nanowires considering all fundamental scattering mechanisms have shown that much of the *ZT* increase comes from the reduced $\kappa_l$.[33] Therefore, in the diffusive limit, we may or may not enjoy advantages in the electronic performance in lower dimensions depending on the details of the scattering processes. This issue deserves further study, but the broad conclusion obtained in Section 3 for ballistic conductors still applies; the packing density of 1D and 2D devices must be high to exceed the absolute power factor of a 3D device, and the individual devices must be small.

Finally, we examine the upper limit performance possible by assuming that $M(E)$ has its ideal shape – a delta function. Mahan and Sofo[9] showed that a delta-shaped $\Xi(E)$ gives the best thermoelectric efficiency because it makes the electronic heat conduction zero, $\kappa_e = \kappa_0 - TS^2G = 0$, which minimizes the denominator of *ZT*. For $M(E) = M_0 \delta(E - E_C)$, we find

$$G_{delta} = \frac{2q^2}{h} M_0 \frac{1}{k_B T} \frac{e^{(E_C - E_F)/k_B T}}{\left(e^{(E_C - E_F)/k_B T} + 1\right)^2} \tag{14a}$$

$$S^2 G_{delta} = \frac{2k_B^2}{h} M_0 \frac{1}{k_B T} \frac{e^{(E_C - E_F)/k_B T}}{\left(e^{(E_C - E_F)/k_B T} + 1\right)^2} \left(\frac{E_C - E_F}{k_B T}\right)^2. \tag{14b}$$

Fig. 5 shows that $S^2 G_{delta}$ has two peaks at $\eta_F \sim \pm 2.4$ and that its maximum value $S^2 G_{delta,max} \sim 2k_B M_0 / hT \times 0.44$ is proportional to $M_0$. To explore the potential benefit from engineering $M(E)$, we compare the power factors calculated from $M_{3D}(E)$ in (11c) and $M_0 \delta(E - E_C)$. We determine the $M_0$ that makes $M_{eff}$ the same for the two cases for $A = 1$ cm$^2$ and then compare the maximum power factors. The result shows that $S^2 G_{3D,max} = 12.6$ W/K$^2$ and $S^2 G_{delta,\max} = 19.3 \text{ W/K}^2$. Therefore, shaping $M_{3D}(E)$ into a delta-function gives a 53 % improvement in power factor. It should be noted, however, that we have assumed that $M_{eff}$ is the same in both cases.

We can also compare the $M_{2D}(E)$ in (11b) and $M_{1D}(E)$ in (11a) with the delta-function. The comparison shows a 23 % improvement over 2D and a 10 % improvement over 1D power factors. The advantage of the delta-function decreases with decreasing dimensionality because the shape of $M(E)$ improves as dimensionality decreases. It should be noted, however, that the magnitude of $M(E)$ is also important as well as the shape of it. Shaping $M(E)$ promises some benefit, but the benefits would also come by increasing $M(E)$. Therefore, it is worth exploring the possibilities of engineering both the shape and the magnitude of $M(E)$ to maximize the thermoelectric efficiency. Molecular thermoelectrics[34,35] may have potential because $M(E)$ is inherently a broadened delta-function, and its magnitude might be greatly increased by connecting many molecules in parallel. In another recent experiment,[5] an increase of *ZT* was reported for bulk PbTe doped by Tl. This is believed to be due to the additional resonant energy level, which improves both the shape and magnitude of $M(E)$.

## 5. Conclusions

In this paper, we examined the role of dimensionality on the electronic performance of thermoelectric devices using the Landauer formalism. We showed that the transmission ($T(E)$) and the number and distribution of conducting channels ($M(E)$) are major factors determining thermoelectric transport coefficients. We also found that the "transport distribution"[9,10] is proportional to the product, $T(E)M(E)$. Assuming ballistic transport ($T(E) = 1$), we were able to show quantitatively how much more efficiently the modes are utilized in 1D than in 2D and 3D. It is hard, however, to realize the advantage because the quantum wires or wells should be closely packed, and their thicknesses should be very small. To first order, these conclusions also apply in the diffusive limit.

Using the Landauer approach, we also discussed the possible benefits from engineering $M(E)$ into a delta-function. For the same effective number of conducting channels, the improvement over a parabolic band in 3D is about 50 %. As dimensionality decreases, the shape of $M(E)$ becomes closer to a delta-function. However, this does not necessarily mean that 1D is better than 2D or 3D because the magnitude of $M(E)$ is also important. It is not dimensionality itself that is important; it is the shape and the magnitude of $M(E)$. We conclude that reduced dimensionality *per se*, does not hold great promise for improving the electronic part of the figure of merit. Engineering bandstructures through size quantization, strain, crystal orientation, etc., should, however, be carefully explored in addition to the efforts to reduce $\kappa_l$ [4,33].

**Acknowledgment**

This work was supported by the National Science Foundation under grant number ECS-0609282. Computational support was provided by the Network for Computational Nanotechnology, supported by the National Science Foundation under cooperative agreement EEC-0634750. It is a pleasure to acknowledge helpful discussions with L. Siddiqui and A. N. M. Zainuddin at Purdue University and Professor Ali Shakouri at the University of California at Santa Cruz.

**References**

1   H. J. Goldsmid, *Thermoelectric Refrigeration* (Plenum Press, New York, 1964).

2   R. Venkatasubramanian, E. Siivola, T. Colpitts, and B. O'Quinn, Nature **413,** 597-602 (2001).

3   A. I. Boukai, Y. Bunimovich, J. Tahir-Kheli, J.-K. Yu, W. A. Goddard Iii, and J. R. Heath, Nature **451,** 168-171 (2008).

4   A. I. Hochbaum, R. Chen, R. D. Delgado, W. Liang, E. C. Garnett, M. Najarian, A. Majumdar, and P. Yang, Nature **451,** 163-167 (2008).

5   J. P. Heremans, V. Jovovic, E. S. Toberer, A. Saramat, K. Kurosaki, A. Charoenphakdee, S. Yamanaka, and G. J. Snyder, Science **321,** 554-557 (2008).

6   L. D. Hicks and M. S. Dresselhaus, Physical Review B **47,** 12727 (1993).

7   L. D. Hicks and M. S. Dresselhaus, Physical Review B **47,** 16631 (1993).

8   N. W. Ashcroft and N. D. Mermin, *Solid State Physics* (Saunders College Publishing, Philadelphia, 1976).

9   G. D. Mahan and J. O. Sofo, Proceedings of the National Academy of Sciences of the United States of America **93,** 7436-7439 (1996).

10  T. J. Scheidemantel, C. Ambrosch-Draxl, T. Thonhauser, J. V. Badding, and J. O. Sofo, Physical Review B **68,** 125210 (2003).

11  A. Shakouri and C. LaBounty, in *Material optimization for heterostructure integrated thermionic coolers*, Baltimore, MD, 1999, p. 35-39.

12  G. Chen and A. Shakouri, Journal of Heat Transfer **124,** 242-252 (2002).

13  T. E. Humphrey and H. Linke, Physical Review Letters **94,** 096601 (2005).

14  R. Landauer, IBM J. Res. Develop. **1,** 223 (1957).

[15] U. Sivan and Y. Imry, Physical Review B **33,** 551 (1986).

[16] P. Streda, Journal of Physics: Condensed Matter **1,** 1025-1027 (1989).

[17] P. N. Butcher, J. Phys.: Condens. Matter **2,** 4869-4878 (1990).

[18] H. v. Houten, L. W. Molenkamp, C. W. J. Beenakker, and C. T. Foxon, Semiconductor Science and Technology **7,** B215-B221 (1992).

[19] L. D. Hicks, T. C. Harman, X. Sun, and M. S. Dresselhaus, Physical Review B **53,** R10493 (1996).

[20] K. Chao and M. Larsson, in *Physics of Zero- and One-Dimensional Nanoscopic Systems*; *Vol. 156* (Springer, Berlin, 2007), p. 151-186.

[21] S. Datta, *Quantum Transport: Atom to Transistor* (Cambridge University Press, New York, 2005).

[22] G. D. Mahan, Journal of Applied Physics **76,** 4362-4366 (1994).

[23] S. Datta, *Electronic Transport in Mesoscopic Systems* (Cambridge University Press, Cambridge, 1995).

[24] L. W. Whitlow and T. Hirano, Journal of Applied Physics **78,** 5460-5466 (1995).

[25] A. Shakouri, Proceedings of the IEEE **94,** 1613-1638 (2006).

[26] M. Lundstrom and J. Guo, *Nanoscale Transistors: Device Physics, Modeling and Simulation* (Springer, New York, 2006).

[27] M. Lundstrom, *Fundamentals of Carrier Transport*, 2nd ed. (Cambridge University Press, Cambridge, 2000).

[28] J. Wang, E. Polizzi, A. Ghosh, S. Datta, and M. Lundstrom, Applied Physics Letters **87,** 043101 (2005).

[29] S. Jin, M. V. Fischetti, and T.-w. Tang, Journal of Applied Physics **102,** 083715 (2007).

[30] R. Kotlyar, B. Obradovic, P. Matagne, M. Stettler, and M. D. Giles, Applied Physics Letters **84,** 5270-5272 (2004).

[31] L. Donetti, F. Gamiz, J. B. Roldan, and A. Godoy, Journal of Applied Physics **100,** 013701 (2006).

[32] D. A. Broido and T. L. Reinecke, Physical Review B **64,** 045324 (2001).

[33] N. Mingo, Applied Physics Letters **84,** 2652-2654 (2004).

[34] M. Paulsson and S. Datta, Physical Review B **67,** 241403 (2003).

[35] P. Reddy, S.-Y. Jang, R. A. Segalman, and A. Majumdar, Science **315,** 1568-1571 (2007).

**Figure Captions**

Fig. 1. Sketches of (a)-(c) the density of states (*D*) and (d)-(f) the number of modes (*M*) for 1D, 2D, and 3D conductors with single parabolic subbands.

Fig. 2. Model calculation (*T* = 300 K) results for the $|S|$ vs. $\eta_F$ for 1D, 2D, and 3D ballistic conductors. For the same $\eta_F$, $|S_{3D}| > |S_{2D}| > |S_{1D}|$. The arrows indicate the magnitude of *S* at $\eta_{F,\max}$ where the power factor in Fig. 3 becomes the maximum in each dimension. We observe $|S_{1D}(\eta_{F,\max})| > |S_{2D}(\eta_{F,\max})| > |S_{3D}(\eta_{F,\max})|$.

Fig. 3. Model calculation ($m^* = m_0$, *T* = 300 K) results for the power factor ($S^2G$) vs. $\eta_F$ for (a) 1D (b) 2D and (c) 3D ballistic conductors. Power factor shows a maximum with $\eta_{F,\max} = -1.14, -0.367$, and 0.668 in 1D, 2D, and 3D, respectively.

Fig. 4. Model calculation ($m^* = m_0$, *T* = 300 K) results for electrical conductance (*G*) vs. $\eta_F$ for (a) 1D (b) 2D and (c) 3D ballistic conductors. It increases more rapidly with $\eta_F$ in 1D than in 2D or 3D.

Fig. 5. Calculation results for the power factor vs. $\eta_F$ for $M(E) = M_0 \delta(E - E_C)$. The expression is shown in (14b). The maximum power factor appears at $\eta_F \sim \pm 2.4$, and the maximum value is $\sim 2k_B M_0 / hT \times 0.44$.

Figure 1

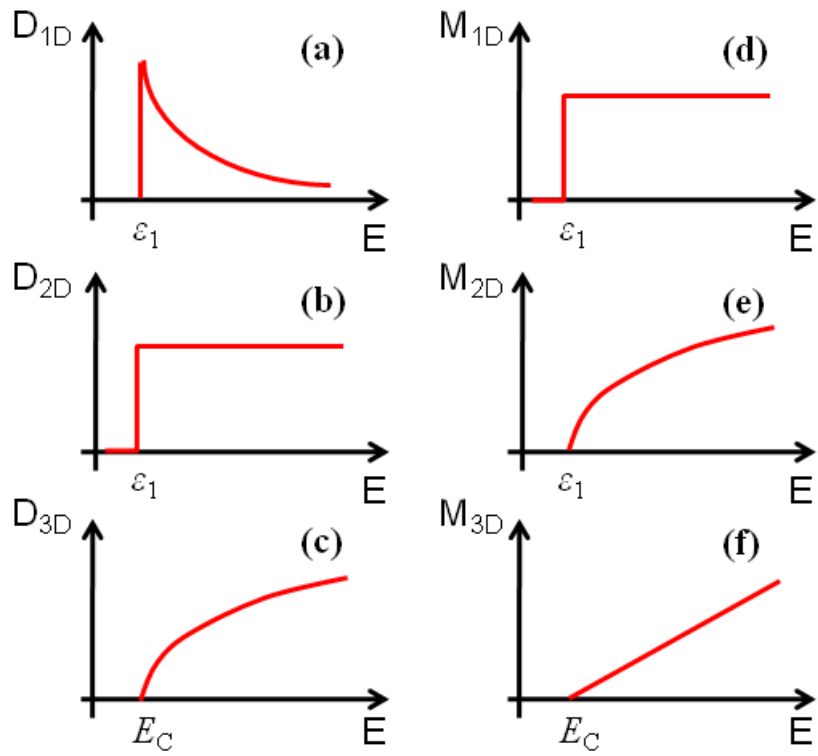

Figure 2

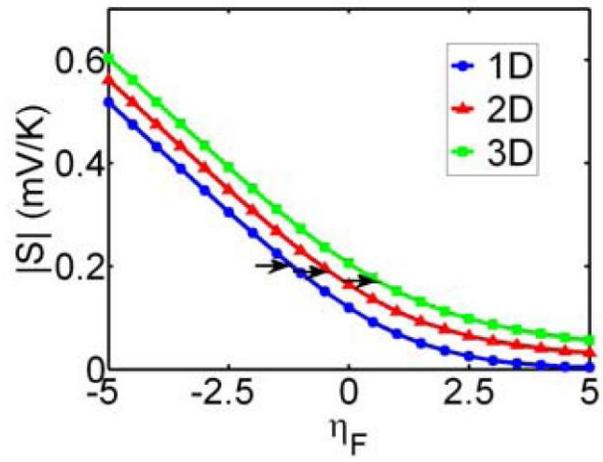

Figure 3

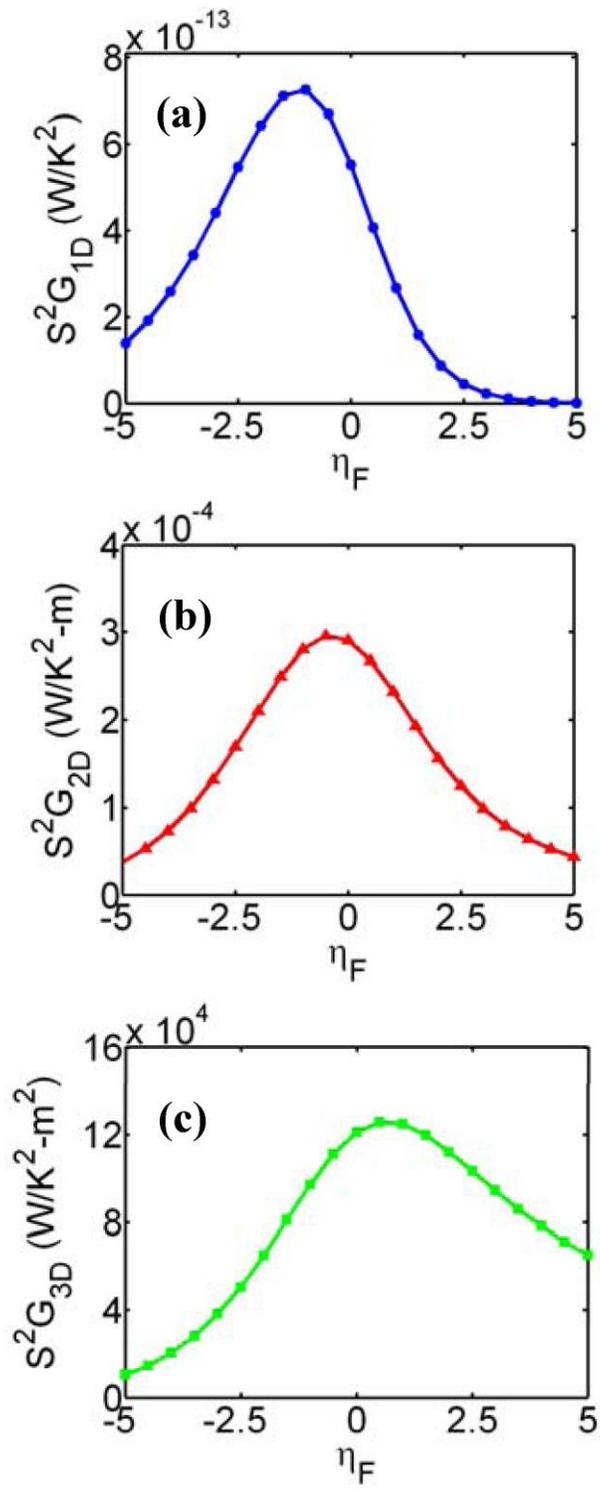

Figure 4

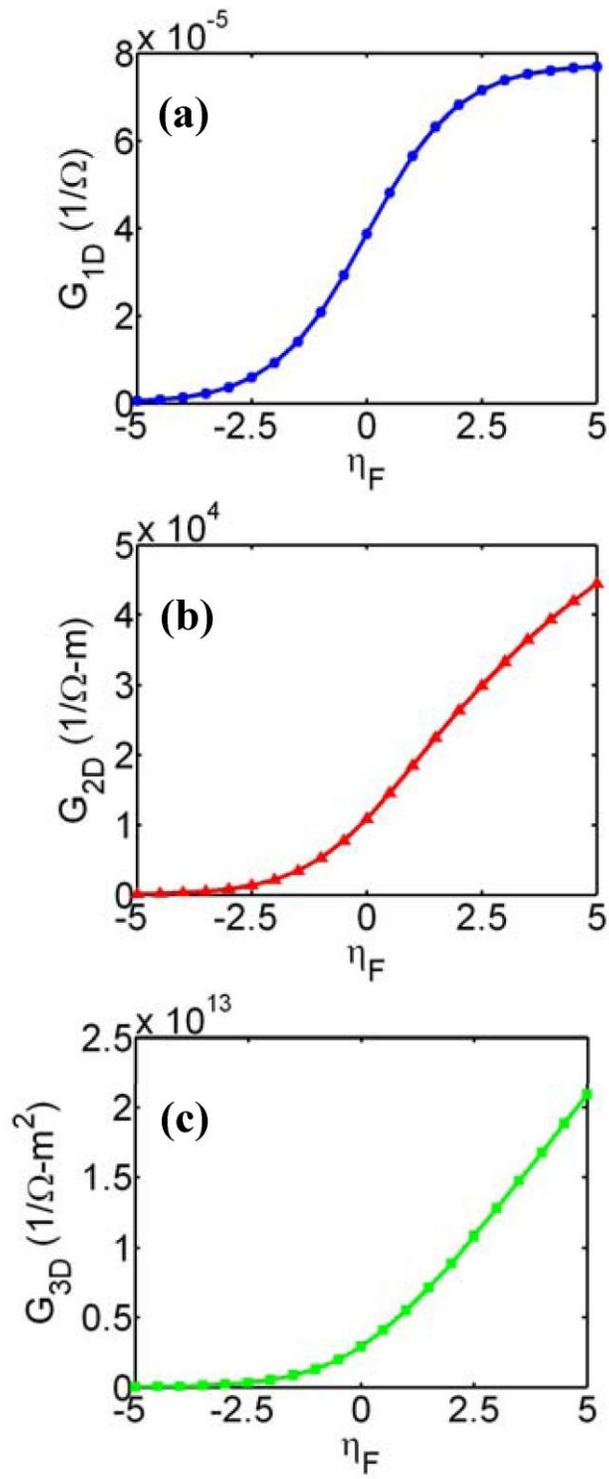

Figure 5

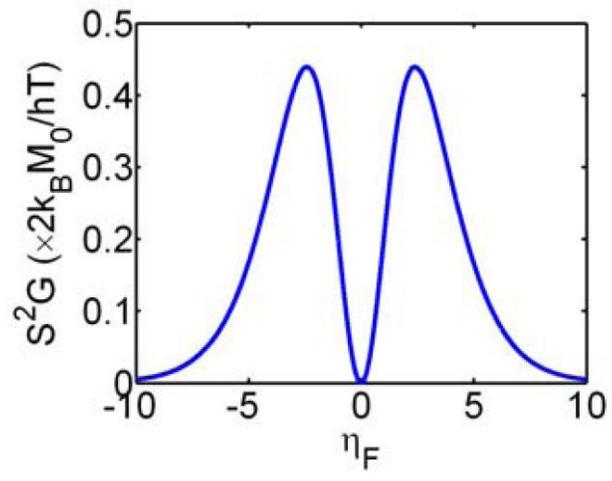